\documentclass[pra,twocolumn,showpacs,nofootinbib,floatfix,superscriptaddress]{revtex4}

\usepackage{amsmath,amsfonts,amssymb,bm}
\usepackage{dcolumn}
\usepackage[final]{graphicx}
\usepackage{bm}

\begin{document}

\title{Optimal quantum control of Bose Einstein condensates in magnetic microtraps}

\author{Ulrich Hohenester}\email{ulrich.hohenester@uni-graz.at}
\author{Per Kristian Rekdal}
\affiliation{Institut f\"ur Physik,
  Karl--Franzens--Universit\"at Graz, Universit\"atsplatz 5,
  8010 Graz, Austria}
  
\author{Alfio Borz\`{\i}}
\affiliation{Institut f\"ur Mathematik,
  Karl--Franzens--Universit\"at Graz, Heinrichstra\ss e 36,
  8010 Graz, Austria}
  
\author{J\"org Schmiedmayer}
\affiliation{Atominstitut der \"Osterreichischen Universit\"aten,
  TU--Wien, Stadionallee 2, 1020 Wien, Austria}

\date{January 15, 2007}

\begin{abstract}

Transport of Bose-Einstein condensates in magnetic microtraps, controllable by external parameters such as wire currents or radio-frequency fields, is studied within the framework of optimal control theory (OCT). We derive from the Gross--Pitaevskii equation the optimality system for the OCT fields that allow to efficiently channel the condensate between given initial and desired states. For a variety of magnetic confinement potentials we study transport and wavefunction splitting of the condensate, and demonstrate that OCT allows to drastically outperfrom more simple schemes for the time variation of the microtrap control parameters.

\end{abstract}

\pacs{03.75.-b,39.20.+q,39.25.+k,02.60.Pn}

%  03.75.-b Matter waves
%  39.20.+q Atom interferometry techniques
%  39.25.+k Atom manipulation
%  02.60.Pn Numerical optimization

\maketitle

 %%%%%%%%%%%%%%%%%%%%%%
 %%%  INTRODUCTION  %%%
 %%%%%%%%%%%%%%%%%%%%%%

\section{Introduction}

Trapping and coherent manipulation of cold neutral atoms in microtraps near surfaces of atomic chips is a promising approach towards full control of matter waves on small scales \cite{folman:00,haensel:01b,folman:02}. This field of atom optics is making rapid progress, driven both by the fundamental interest in quantum systems and by the prospect of new devices based on quantum manipulations of neutral atoms. Lithographic and other surface-patterning processes nowadays allow to build complex atom chips which combine many traps, waveguides, and other elements, in order to realize controllable composite quantum systems \cite{zoller:02} as needed, e.g., for the implementation of quantum information devices \cite{nielsen:00}. Such microstructured surfaces have been highly successful and form the basis of a growing number of experiments \cite{hommelhof:05}.

The possibility to store, manipulate \cite{guenther:05,luo:04,krueger:03,schumm:05,hofferberth:06,wang:05}, and measure a single quantum system with extremely high precision has initiated great stimulus in various fields of research, ranging from atom interferometry \cite{haensel:01,andersson:02,wang:05,schumm:05,jo:06}, over quantum gates \cite{calarco:00,charron:06,treutlein:06} and resonant condensate transport \cite{paul:05}, to microscopic magnetic-field imaging \cite{wildermuth:05}. In the vast majority of these schemes the wavefunction of the Bose-Einstein condensate, trapped in the vicinity of an atom chip, is manipulated through variation of the magnetic confinement potential. This is achieved by changing the currents through the gate wires mounted on the chip or modifying the strength of additional radio-frequency fields \cite{folman:02,lesanovsky:06,lesanovsky:06b,hofferberth:06,wildermuth:06}. These external, time-dependent parameters thus provide a versatile control for wavefunction manipulations, and make atom chips attractive candidates for quantum control applications. 

Consider the situation where one is aiming for an efficient wavefunction transfer from a given initial to a final desired state, possibly by passage through a series of other states, or for a conditional quantum gate where atoms have to interact with each other in a well-defined manner. Here the question arises: how should one modify the control fields in order to achieve a most efficient transfer or coupling? This problem was first tackled by H\"ansel et al. \cite{haensel:01} for a trapped-atom inteferometer setup where a dilute condensate should be split through variation of the confinement potential from a single to a double well, such that it ends up in the groundstate of the final double well potential. These authors devised a scheme that optimizes adiabatic transfer by minimizing transitions to excited states.

In this paper we study quantum control of Bose-Einstein condensates in magnetic microtraps within the framework of optimal control theory (OCT). Here, the objective of the control is quantified through a cost function, which is then minimized subject to the condition that the time dynamics of the condensate is governed by the Gross--Pitaevskii equation \cite{dalfovo:99,leggett:01}. We will show that optimal control theory provides a versatile tool for determining efficient control strategies, and is applicable for realistic confinement potentials, one- and two-dimensional problems, and nonlinearities in the condensate dynamics. Optimal control theory is a mathematical device that allows for a general determination of efficient control strategies \cite{peirce:88,borzi.pra:02}, and has found widespread applications for, e.g., molecules \cite{rabitz:00,tesch:02}, atoms \cite{calarco:04,koch:04}, or semiconductors \cite{hohenester.prl:04}. We believe that there are a number of reasons that render OCT ideal for quantum control of condensates in atom chips. First, it is only the cost function that determines the optimal control. There is no additional bias, such as, e.g., in the adiabatic scheme where scattering losses are minimized throughout the whole transfer process, and consequently OCT allows to explore a larger portion of the control space. In addition no knowledge of the stationary solutions of the  Gross-Pitaevskii equation is required in OCT, contrary to the adiabatic scheme where ground and excited states must be determined for every control configuration. Since the optimal control corresponds to a minimum in the control space, the solutions are robust with respect to small fluctuations of the external parameters, which can never be avoided in real experiments. Finally, decoherence effects, that also play a role in atom chips \cite{henkel:99,folman:02,scheel:05,skagerstam:06}, can be naturally incorporated into OCT calculations \cite{hohenester.prl:04,jirari:05,hohenester.prb:06}.

We have organized our paper as follows. In sec.~\ref{sec:theory} we introduce to the realm of optimal quantum control, and derive the optimality system for condensate transport in atom chips. In sec.~\ref{sec:results} we present results for condensate splitting in simple and realistic confinement potentials. We demonstrate that our scheme is applicable for effective one- and two-dimensional geometries, and for nonlinearities in the condensate transport. Finally, in sec.~\ref{sec:summary} we summarize and draw some conclusions.

 %%%%%%%%%%%%%%%%
 %%%  Theory  %%%
 %%%%%%%%%%%%%%%%

\section{Theory}\label{sec:theory}

We consider a coherent ensemble of Bose-Einstein condensed atoms confined in a potential $V(\bm r,\lambda(t))$ produced by a magnetic microtrap. $\lambda(t)$ is a control parameter that describes the variation of the confining potential when changing the external parameters, such as currents through the microtrap wires or frequency and strength of additional radio-frequency fields \cite{haensel:01,folman:02,lesanovsky:06} (for details see below). Through $\lambda(t)$ it is possible to manipulate the Bose-Einstein condensate, e.g., to split and reunite it by varying the potential from a single to a double well and vice versa. We assume that $\lambda(t)$ is a single-valued, real parameter, although different situations, e.g., microtraps controlled by several parameters, could be treated equally well. In the following we shall assume for simplicity that $\lambda(t)$ can only take values between zero and one. The mean-field dynamics of the condensate is described by the Gross--Pitaevskii equation \cite{dalfovo:99,leggett:01}
\begin{equation}\label{eq:schroedinger}
  i\dot\psi(\bm r,t)=\left(-\frac 1 2 \nabla^2+V(\bm r,\lambda(t))+
  g\left|\psi(\bm r,t)\right|^2\right)\psi(\bm r,t)\,,
\end{equation}
with $g$ a coupling constant related to the scattering length of the atoms.\footnote{We set $\hbar=1$, and measure mass in units of the atom mass and length in units of micrometers. For $^{87}$Rb atoms the time and energy scales are then given by 1.37 milliseconds and 5.58 nano Kelvins, respectively.} Suppose that initially the system is in the groundstate $\psi_0$ for the potential $V(\bm r,\lambda=0)$. Upon varying $\lambda(t)$ in the time interval $t\in[0,T]$ from zero to one, the system will pass through a sequence of states and will end up in the final state $\psi(T)$. Within the field of quantum control one is usually seeking for an optimized time evolution of $\lambda(t)$ that allows to channel the system from the initial state $\psi_0$ at time zero to a desired state $\psi_d$ at final time $T$. In accordance to Ref.~\cite{haensel:01}, we assume $\psi_d$ to be the groundstate for the potential $V(\bm r,\lambda=1)$ at time $T$. 
Let
\begin{equation}\label{eq:cost}
  J(\psi,\lambda)=\frac 1 2 \bigl(1-\bigl|\langle \psi_d|\psi(T)\rangle\bigr|^2\bigr)
  +\frac\gamma 2\int_0^T \left(\dot\lambda(t)\right)^2\,dt
\end{equation}
be the cost function that rates how good the final state $\psi(T)$ matches the desired state $\psi_d$, with $\langle u|v\rangle=\int d\bm r\,u^*(\bm r)v(\bm r)$ the usual inner product.\footnote{We assume that the wavfunction $\psi(\bm r,t)$ is normalized to one. In comparison to, e.g., ref.~\cite{dalfovo:99}, where the wavefunction is normalized to the number of atoms $N$ in the condensate, the nonlinearity parameter $g$ in eq.~\eqref{eq:schroedinger} is therefore assumed to incorporate $N$.} The first term on the right-hand side becomes zero for $\psi(T)=\psi_d$ and maximal if final and desired state do not overlap.\footnote{In contrast to the $\frac 1 2 \|\psi(T)-\psi_d\|^2$ cost function used in ref.~\cite{borzi.pra:02}, in expression~\eqref{eq:cost} the final wavefunction has to match the desired one only up to a global phase $e^{i\phi}$. This allows the substraction of constant values in the confinement potential $V(\bm r,\lambda)$, as discussed in more detail in appendix~\ref{app:phase}, which proves particularly useful for magnetic confinement potentials with large energy offsets \cite{lesanovsky:06}.} The second term on the right-hand side favours control fields $\lambda(t)$ with a smooth time variation and is needed to make the quantum control problem well posed. $\gamma$ is a weighting parameter that determines the importance of the two different control strategies of wavefunction matching and smooth control fields. We shall use small $\gamma$ values throughout, such that the cost function $J(\psi,\lambda)$ is dominated by the first term. The control problem under consideration thus becomes the minimization of the cost function $J(\psi,\lambda)$ subject to the condition that $\psi(t)$ fulfills the Gross--Pitaevskii equation \eqref{eq:schroedinger}.

Within the field of optimal control theory (OCT) one uses Lagrange multipliers to turn this constrained minimization problem into an unconstrained one. For this purpose we define the Lagrange function
\begin{eqnarray}\label{eq:lagrange}
  &&L(\psi,p,\lambda)=J(\psi,\lambda)\nonumber\\&&\quad+
  \Re e\left<p,i\dot\psi-\left(-\frac 1 2\nabla^2+V_\lambda+g|\psi|^2\right)\psi\right>\,,
\end{eqnarray}
with the abbreviation $\left<u,v\right>=\int_0^T dt\int d\bm r\,u^*v$, and $p(t)$ the Lagrange multiplier. We next utilize the fact that the Lagrange function has a saddle point at the minimum of $J(\psi,\lambda)$, i.e. all three derivatives $\delta L/\delta\psi$, $\delta L/\delta p$, and $\delta L/\delta\lambda$ must be zero. Performing usual functional derivatives in eq.~\eqref{eq:lagrange} we obtain after some variational calculation the following optimality system
\begin{subequations}\label{eq:optimality}
\begin{eqnarray}
&&i\dot\psi=\left(-\frac 1 2\nabla^2+V_\lambda+g|\psi|^2\right)\psi
\label{eq:oct.forward}\\
&&i\dot p=\left(-\frac 1 2\nabla^2+V_\lambda+2g|\psi|^2\right)p+
g\,\psi^2\,p^*
\qquad\label{eq:oct.backward}\\
&&\gamma\ddot\lambda=-\Re e\langle\psi|\frac{\partial V_\lambda}{\partial\lambda}|p\rangle
\label{eq:control}\,,
\end{eqnarray}
\end{subequations}
which has to be solved together with the initial and terminal conditions 
\begin{subequations}
\begin{eqnarray}
  &&\phantom{i}\psi(0)=\psi_0\label{eq:bc.forward}\\
  &&ip(T)=-\langle\psi_d|\psi(T)\rangle\,\psi_d\label{eq:bc.backward}\\
  &&\phantom{i}\lambda(0)=0\,,\quad\lambda(T)=1\,.\label{eq:bc.control}
\end{eqnarray}
\end{subequations}
The right-hand side of eq.~\eqref{eq:bc.backward} follows from the functional derivative $\delta J/\delta\psi$. Notice that while the state equation \eqref{eq:oct.forward} with initial condition $\psi(0)=\psi_0$ evolves forward in time, the adjoint equation \eqref{eq:oct.backward} with terminal condition \eqref{eq:bc.backward} is marching backwards.~\footnote{For the optimal control field $\lambda(t)$ the adjoint equation \eqref{eq:oct.backward} describes the fluctuations of the system around $\psi(t)$. For this reason it is often referred to as the {\em sensitivity equation}.\/ Incidentally, eq.~\eqref{eq:oct.backward} for the adjoint variable $p$ closely resembles the time-dependent Bogoliubov--de Gennes equation \cite{leggett:01} which usually serves as a starting point for the treatment of collective excitations.} The control equation \eqref{eq:control} determines the optimal control.

In most cases of interest one is not able to directly guess $\lambda(t)$ such that eqs.~(\ref{eq:oct.forward}--c) are simultaneously fulfilled, and one has to employ an iterative scheme. In this work we follow ref.~\cite{borzi.pra:02} and formulate a numerical algorithm that solves the optimality system (\ref{eq:oct.forward}--c) for given initial and desired configurations $\psi_0$ and $\psi_d$, respectively. To solve this problem, we apply a gradient-type minimization algorithm, which, starting from an initial guess for $\lambda(t)$, determines a search direction for an improved control. We first solve eq.~\eqref{eq:oct.forward} with initial condition $\psi(0)=\psi_0$ forwards in time. Once the wavefunction $\psi(T)$ at time $T$ is computed, the final condition for $p(T)$ can be calculated from eq.~\eqref{eq:bc.backward} and the adjoint equation of motion \eqref{eq:oct.backward} is solved backwards in time. The gradient of $L$ with respect to $\lambda$ becomes
\begin{equation}\label{eq:grad.L}
  \frac{\delta L}{\delta\lambda}=-\gamma\ddot\lambda-\Re e\langle\psi|
  \frac{\partial V_\lambda}{\partial\lambda}|p\rangle\,,
\end{equation}
which gives the search direction for an improved control that minimizes $J(\psi,\lambda)$. In the following we employ for the minimum search either the usual gradient method or the BFGS quasi-Newton method \cite{bertsekas:99,borzi.pra:02}. Details of our solution scheme for the Schr\"odinger-type equations are given in appendix \ref{app:details}.

We emphasize that the choice of our cost function \eqref{eq:cost} is by no means unique. For instance, one could add an additional $\eta\left<\psi,\left(-\nabla^2/2+ V_\lambda+g|\psi|^2\right)\psi\right>$ term, with another weighting parameter $\eta$, to minimize the total energy within the transfer process and to favour adiabatic processes. Another possibility would be to make $\gamma$ in eq.~\eqref{eq:cost} time dependent and to penalize control variations more strongly at the beginning and end of the transfer process, such that $\lambda(t)$ is turned on and off sufficiently smooth, which might be beneficial for experimental implementations. Alternatively, through a slight variant of the cost function the system can be forced to pass through a number of desired states \cite{borzi.pra:02} or to acquire a certain phase~\cite{charron:06}.
One could also use a slight variant of our approach to obtain an optimization scheme for spatial geometries of waveguides and two-arm interferometers through which a condensate can propagate without creating excitations, as briefly outlined in appendix \ref{app:waveguide}.

\section{Results}\label{sec:results}

\begin{figure}
\centerline{\includegraphics[width=0.85\columnwidth]{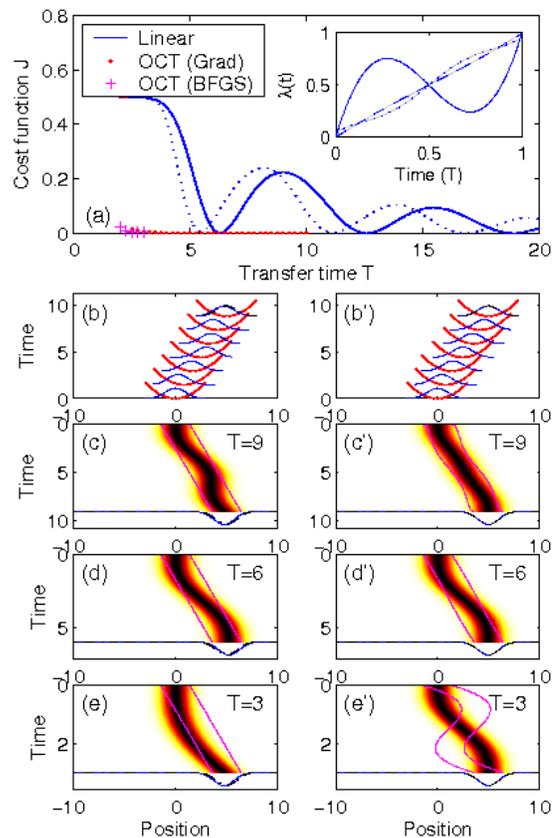}}
\caption{
(Color online) Results of our simulation for the single-well potential \eqref{eq:vsingle} and $x_0=5$. (a) Cost function $J(\psi,\lambda)$ for different transfer times $T$ and for a linear time variation of $\lambda$ (solid line) and optimized variations (symbols). The dotted line shows results of a simulation where an additional $x^4$ term is added to the potential (see text). The inset reports the optimized control fields $\lambda(t)$ for transfer times of 3 (solid line), 6 (dashed line), and 9 (dotted line). (b) Time evolution of potential and wavefunction $|\psi(x,t)|$ for linear $\lambda$ and T=9. (c--e) Density plot of $|\psi(x,t)|$ for linear $\lambda$ as a function of time and position, and for transfer times of (c) T=9, (d) T=6, and (e) T=3. On the bottom of each panel we show the desired wavefunction (dashed-dotted line) and $|\psi(x,T)|$ (solid line). The solid lines in the density plots represent the equipotential lines of the confinement potential. (c'--e') Same as (c--e) but for optimized control fields.
}\label{fig:single}
\end{figure}

\begin{figure}
\centerline{\includegraphics[width=0.9\columnwidth]{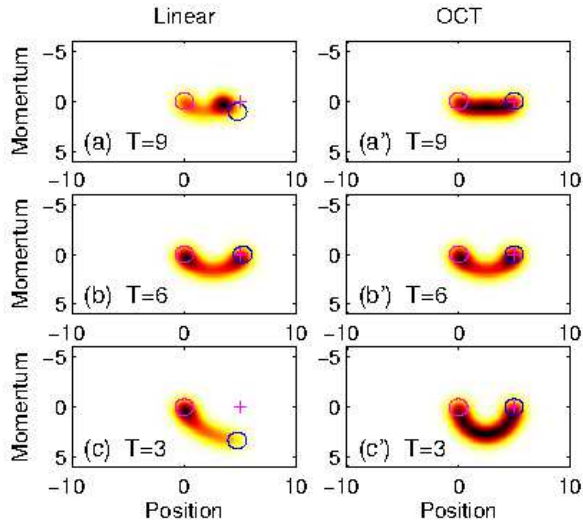}}
\caption{
(Color online) Time-integrated Wigner function $w(x,p)$ for the single-well transfer processes shown in fig.~\ref{fig:single}. The solid lines show the equipotential lines for the Wigner functions of the initial and final wavefunctions $\psi_0$ and $\psi(T)$, respectively.
}\label{fig:wignersingle}
\end{figure}

We first consider the more simple scenario of single atoms or dilute condensates within microtraps, and neglect the nonlinear terms in eqs.~\eqref{eq:oct.forward} and \eqref{eq:oct.backward} by setting $g=0$. The influence of nonlinearities will be discussed at the end. In our optimal quantum control calculations we set $\gamma=10^{-3}$ (smaller values of $\gamma$ turned out to have no noticeable influence on the results) and terminate the optimization loop after several tens to hundreds iterations when the gradient \eqref{eq:grad.L} has become sufficiently small.

We shall consider the scenario where a Bose-Einstein condensate is split into two parts through smooth variation of the magnetic confinement potential from a single to a double well \cite{haensel:01,schumm:05}. In secs.~\ref{sec:single} and B we simulate condensate transport through parabolic-like confinement potentials. These simplified case studies will allow us to grasp the essentials of our optimal control calculations. Transport through realistic magnetic confinement potentials is discussed in sec.~\ref{sec:haensel} and D. Finally, in sec.~\ref{sec:nonlinear} we investigate the influence of nonlinearities in the Gross--Pitaevskii equation.

\subsection{Single well}\label{sec:single}

After successful splitting of the condensate the atoms in the two respective wells can be further transported by shifting the location of the minima. In our first example we will study such transport inside a single well. We will make the assumption that the confinement along $y$ and $z$ is much stronger than along $x$, such that only the dynamics in $x$ direction has to be considered. We assume a potential of the form
\begin{equation}\label{eq:vsingle}
  V(x,\lambda)=\frac 1 2 \left(x-\lambda\,x_0\right)^2\,,
\end{equation}
which has its minimum at $\lambda\,x_0$. By varying within the time interval $[0,T]$ the control parameter from zero to one, the potential minimum becomes shifted from zero to $x_0$. Our objective now is to seek for a time variation $\lambda(t)$ that brings the system from the groundstate $\psi_0$ of the harmonic oscillator centered at $x=0$ to the desired groundstate $\psi_d$ of the displaced harmonic oscillator centered at $x_0$. Although the above model allows under quite general conditions for an analytic solution, see, e.g., Ref.~\cite{caves:80}, the following analysis will turn out to be helpful when discussing the more complicated situation of condensate splitting.

Let us first consider a linear variation $\lambda(t)=t/T$. Figs.~\ref{fig:single}(b--e) report results of our simulations for three selected tranfer times $T$. Panel (b) shows for $T=9$ the modulus of the wavefunction together with the confinement potential at different times, and panel (c) shows a density plot for $|\psi(x,t)|$ of the same transfer process. At the bottom of panel (c) we also plot the final wavefunction (solid line) which somewhat differs from the desired one (dashed-dotted line). Similar behavior is observed for (d) $T=6$ and (e) $T=3$. Finally, in panel (a) we report for the linear $\lambda$-variation the cost function \eqref{eq:cost} as a function of $T$ (solid line), which is high for small values of $T$ and shows an oscillatory behavior with decreasing amplitude for longer transfer times $T$. The decreasing amplitude is due to the fact that with increasing $T$ the time variation of potential \eqref{eq:vsingle} becomes slower, and the system can follow almost adiabatically. 

The oscillations in the cost function are due to the oscillations of the wavefunction inside the single-well potential. To understand their origin, consider the extreme case where the position of the potential minimum is abruptly moved to $x_0$ at time zero, and the system is brought into a highly excited state where the groundstate wavefunction of the harmonic oscillator is displaced by $x_0$ with respect to the new minimum of $V(x,\lambda=1)$. Such displaced groundstates of the harmonic oscillator are known as {\em coherent states}\/ \cite{scully:97} and have a dynamics reminiscent of classical oscillators. As time goes on, the system will start to oscillate with amplitude $x_0$ around its new equilibrium position. Also the $\lambda(t)$ variation with finite speed can be described in terms of such coherent states, as evidenced by the fact that in figs.~\ref{fig:single}(b--e) only the position but not the shape of the wavepacket changes with time.\footnote{This situation corresponds to the classical analogon of a particle attached to a spring which is initially fixed at the origin, and the point where it is fixed is moved to a different position $x_0$ at later time. Only for certain transfer times $T$ the particle will end up in rest position and with no force acting upon it. These times correspond to the minima of the cost function in fig.~\ref{fig:single}(a).} To inquire more into this evolution we shall analyze the Wigner function \cite{zurek:03}
\begin{equation}\label{eq:wigner}
  w(x,p;t)=\int e^{-ips}\psi(x+\frac s 2,t)\psi^*(x-\frac s 2,t)\,ds\,
\end{equation}
for the wavefunction, which is a mixed position-momentum distribution that has many, albeit not all, properties of a classical distribution function. The solid lines in fig.~\ref{fig:wignersingle} show contour lines of the Wigner functions for the initial and final wavefunctions $\psi_0(x)$ and $\psi(x,T)$, respectively. These coherent states are minimum uncertainty states with $\Delta x\,\Delta p=\frac 1 2$. The density plots in the different panels of the figure show the time integrated Wigner function $\int_0^T w(x,p;t)\,dt$ which provides information about the trajectory in phase space. At short transfer times, fig.~\ref{fig:wignersingle}(c), the system ends up in a state that is located close to $x_0$ but with a high momentum. Thus, when the control parameter is kept fixed to $\lambda=1$ at times beyond $T$, the system will continue to oscillate around its new equilibrium position. This final state differs substantially from the desired groundstate $\psi_d$ of the displaced oscillator, and consequently has a rather high cost $J$ [see also fig.~\ref{fig:single}(a)]. With increasing $T$ [figs.~\ref{fig:wignersingle}(a,b)] the momentum of the coherent state decreases and thus $J$ becomes smaller.

\begin{figure}
\centerline{\includegraphics[width=0.85\columnwidth]{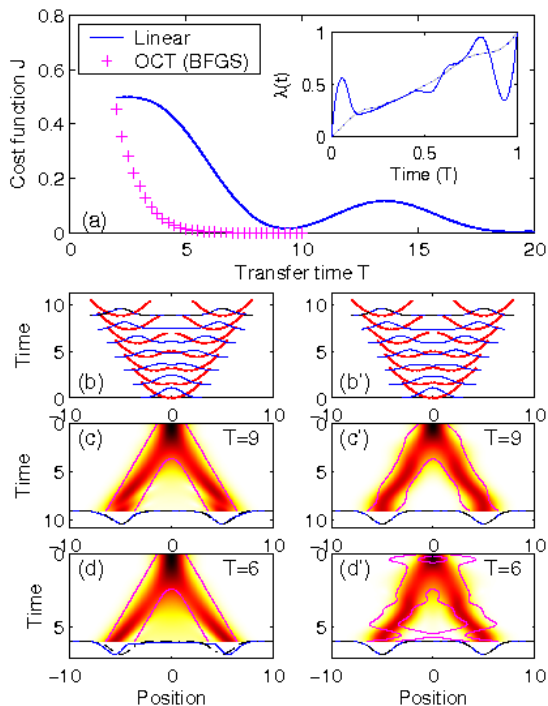}}
\caption{(Color online) Same as fig.~\ref{fig:single} but for the double-well potential \eqref{eq:vdouble}. The inset in panel (a) shows the optimized $\lambda(t)$ for $T=6$ (solid line) and $T=9$ (dashed line). (b) Time evolution of $|\psi(x,t)|$ and $V(x,\lambda(t))$ for $T=9$. (c,d) Density plot of $|\psi(x,t)|$ for (c) $T=9$ and (d) $T=6$. (b'--d') Same as (b--d) but for optimized control fields.
}\label{fig:double}
\end{figure}

We next turn to our optimal quantum control calculations. Here we start from the linear $\lambda(t)=t/T$ function as a guess for the control, and successively improve $\lambda(t)$ according to the scheme described in the previous section. The symbols in fig.~\ref{fig:single}(a) report the cost function for the optimized control fields: throughout $J$ can be drastically improved with respect to the linear $\lambda$ variation. In particular for transfer times beyond say $T=3$ the final wavefunctions perfectly match the desired one. Figs.~\ref{fig:single}(b'--e') report the wavefunction evolution for the optimized control fields depicted in the inset of panel (a). For $T=6$ and $T=9$ the fields deviate only little from the linear dependence, and minor to moderate corrections of $\lambda(t)$ suffice to finally bring the system at $x_0$ to rest. This is also apparent from the Wigner functions shown in fig.~\ref{fig:wignersingle} where the final state (solid contour line) is centered at $x_0$ and has zero momentum.\footnote{It is worth noting that our cost function \eqref{eq:cost} is only governed by the final wavefunction $\psi(T)$, and consequently no guidance of the intermediate wavefunction trajectory is present. Thus, if the linear control fields already work successfully, such as for $T=6$ in fig.~\ref{fig:single}(d), the optimized $\lambda(t)$ and the corresponding transfer process are practically not altered, whereas somewhat stronger deviations can be observed for $T=9$ in fig.~\ref{fig:single}(c). We also emphasize again that the parabolic confinement potential \eqref{eq:vsingle} is special in the sense that it can only change the position but not the shape of the initial wavepacket, and there thus exists a huge variety of different successful control stratgies.} For $T=3$ the control strategy shown in panel (e') becomes noticeably modified: at early times the center of the parabolic confinement potential is quickly shifted and the system is put into a highly nonequilibrium state where it starts to oscillate from left to right. Once it decelerates and reaches the right turning point, the position of the potential minimum is further shifted and the system becomes frozen in the groundstate of the shifted harmonic oscillator. Note that only the quasi-Newton BFGS method is capable of coming up with such control, whereas the more simple gradient scheme cannot and is trapped in a sub-optimal extremum. 

We emphasize that all optimal transport processes discussed here rely on non-equilibrium coherent states, and the resulting transfer processes strongly differ from adiabatic schemes. Finally, in panel (a) the dotted line reports that similar behavior is also found when an additional nonlinear potential term $\eta(x-\lambda x_0)^4/4$ is added, with $\eta=0.2$ in the figure. Also in this case optimal control theory gives control fields (not shown) that allow, in contrast to the linear $\lambda$ variation, perfect transport.

\subsection{Double well}\label{sec:double}

We next turn to the more complicated situation of wavefunction splitting. As a preliminary case study we consider the confinement potential
\begin{equation}\label{eq:vdouble}
  V(x,\lambda)=\begin{cases} 
    \frac 12\left(|x|-\frac {\lambda d} 2\right)^2 & 
      \text{for $|x|>\frac{\lambda d} 4$} \\ & \\
    \frac 12\left(\frac{(\lambda d)^2}8-x^2\right) & \text{otherwise,} \\
  \end{cases}
\end{equation}
which changes from a single-well potential for $\lambda=0$ to a double-well potential with interwell distance $d$ for $\lambda=1$. Potential \eqref{eq:vdouble} is constructed such that it is continous and smooth. Figs.~\ref{fig:double}(b,c) show results for a wavefunction splitting for $T=9$ and for a linear $\lambda(t)$ dependence. The wavefunction becomes split in the first stage of the time evolution, and is transported in the respective minima in the second stage of the transport process. Contrary to the single-well transport, in this second stage also excited vibrational states of the harmonic oscillator that were populated during the initial splitting process are involved, as apparent from the varying shape of $|\psi(x,t)|$ in the density plot of panel (c). Even more striking, the wavefunction shown in panel (d) for the fast transfer process with $T=6$ is split only incompletely, and part of the population remains localized in-between the two wells. Correspondingly, the overlap with the desired groundstate $\psi_d$ of the $V(x,\lambda=1)$ potential is rather poor and the cost function shown in fig.~\ref{fig:double}(a) (solid line) is high for small values of $T$. With increasing $T$ the cost function again exhibits an oscillatory behavior with decreasing amplitude, indicating the onset of adiabatic transport. However, in contrast to the single-well case $J$ keeps a finite value at its minima, which is due to the population of excited vibronic states during splitting and the resulting lack of complete overlap with $\psi_d$.

\begin{figure}
\centerline{\includegraphics[width=0.9\columnwidth]{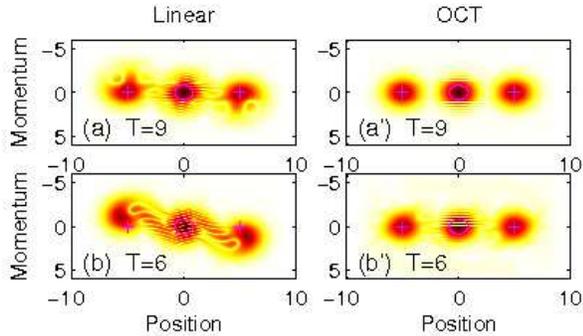}}
\caption{(Color online) Absolute value of Wigner function $|w(x,p;T)|$ at the end of the transfer process [figs.~\ref{fig:double}(b--d)] for linear $\lambda$ variation and for transfer times of (a) $T=9$ and (b) $T=6$. (a',b') Same as (a,b) but for optimized control fields. The interference pattern around the origin is a measure of the coherence properties of the split condensate.
}\label{fig:wignerdouble}
\end{figure}

The symbols in fig.~\ref{fig:double}(a) report the cost function for the optimized process of wavefunction splitting. At short transfer times, say below $T=5$, the optimized control strategies perform significantly better than the linear ones, but the overlap with the desired state is not perfect. We emphasize that these results do not exclude the possibility of more efficient transfer in regions of the control space that were not explored by our minimization scheme. For transfer times beyond $T=6$ the cost function drops below a value of $10^{-3}$ indicating the onset of efficient wavefunction splitting. Figs.~\ref{fig:double}(b',c') show the wavefunction evolution for $T=9$, which is not drastically altered in comparison to the evolution for the linear scheme. A slight modification of the control function $\lambda(t)$ suffices to channel the system to the desired state at time $T$. Fig.~\ref{fig:wignerdouble} shows the Wigner functions $w(x,p;T)$ at the end of the transfer processes. For the optimized control shown in panel (a') it consists of two coherent-state features at the positions of the two minima of the double well potential (see cross symbols), indicating that $\psi(T)$ matches the respective single-well groundstates, and an interference pattern at position zero due to the superposition nature of the wavefunction $\psi(T)$  \cite{zurek:03}. In contrast, the Wigner function for the linear time evolution shown in panel (a) exhibits an asymmetric shape at the positions of the potential minima, that can be traced back to the superposition of ground and excited vibronic states within the respective minima. For the short transfer time of $T=6$ the optimized $\lambda(t)$ shown in the inset of fig.~\ref{fig:double}(a) substantially differs from a linear behavior. As apparent from the corresponding wavefunction evolution shown in fig.~\ref{fig:double}(d'), at early times the potential is quickly transformed from a single to a double well, and the system is thereby brought into a highly excited state. Similar to the fast single-well transport described above, such states can be manipulated and transported on shorter timescales. Indeed, the final stage of the transfer process is reminiscent of the final stage of wavefunction transport shown in fig.~\ref{fig:single}(e').

\subsection{Magnetic confinement of H\"ansel et al.}\label{sec:haensel}

\begin{figure}
\centerline{\includegraphics[width=0.85\columnwidth]{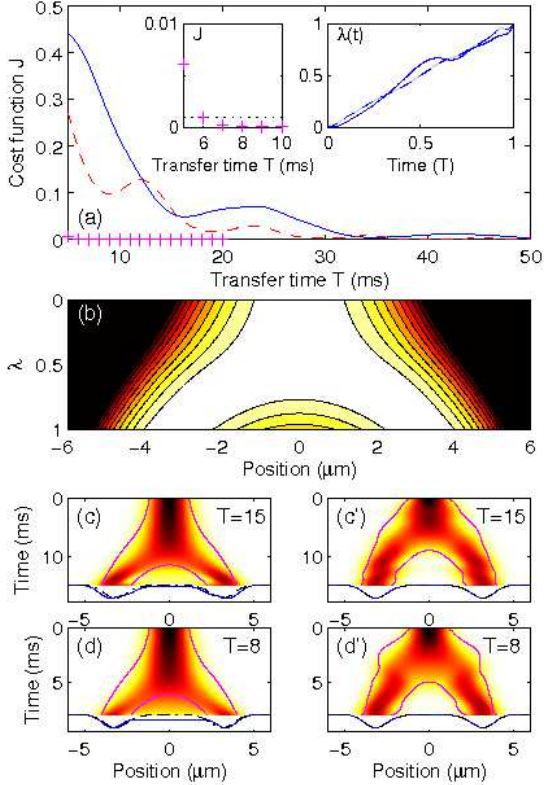}}
\caption{(Color online)
Results for the magnetic microtrap of H\"ansel {\em et al.}\/ \cite{haensel:01} and for ${}^{87}$Rb atoms confined in the $m_F=2$ state. (a) Cost function for linear $t/T$ (solid line), square-root $\sqrt{t/T}$ (dashed line), and optimized (symbols) variation of $\lambda(t)$. The optimized $J(\psi,\lambda)$ is magnified in the left inset. The right inset reports the optimal $\lambda(t)$ for $T=8$ ms (solid line) and $T=15$ ms (dashed line). (b) Contour plot of magnetic confinement potential as a function of $\lambda$. (c,d) Wavefunction evolution for linear variation of $\lambda$ and transfer times of (c) $T=15$ ms and (d) $T=8$ ms. (c',d') Same as (c,d) but for optimized control.
}\label{fig:haensch1d}
\end{figure}

In H\"ansel {\em et al.}\/ \cite{haensel:01} the authors studied wavefunction splitting for a realistic magnetic microtrap. They devised a control scheme that favours adiabatic transport by minimizing, throughout the whole transfer process, excitations to excited states, and demonstrated that such approach can perform significantly better in comparison to more simplified control strategies. In this section we re-examine their scheme within the framework of optimal control theory. We use the same model parameters for the magnetic microtrap\footnote{The strength of the field component $B_{0,y}$ should be 30 G, rather than the 20 G given in eq.~(2) of ref.~\cite{haensel:01}, in order to match the distance of 35 $\mu$m between trap and surface.} where confinement along $x$ is provided by three parallel wires oriented along the $y$ direction, with an inter-wire distance of 20 $\mu$m. The current $I_{\rm ext}$ through the central wire is opposite to the currents $I_c$ through the outer wires. Introducing a current modulation by means of the control parameter $\lambda$ via
\begin{eqnarray}
  I_{\rm ext} &=& 140\phantom{.}+\lambda\times 2.91 \,\,\text{mA}\nonumber\\
  I_{c} &=& 0.25 + \lambda\times 4.4\phantom{0} \,\,\text{mA}\,
\end{eqnarray}
produces a magnetic confinement along $x$ that changes from a single well at $\lambda=0$ to a double well at $\lambda=1$, as shown in fig.~\ref{fig:haensch1d}(b). For the linear variation of $\lambda(t)$ wavefunction splitting is shown in panels (c) and (d) for transfer times of 15 and 8 ms, respectively. In both cases the splitting is too fast to allow the system to become localized in the two mimima of the double well, and a significant portion of the population remains in-between the two wells. This is also apparent from the cost function shown in fig.~\ref{fig:haensch1d}(a) (solid line) that reports large $J$ values over a wide range of transfer times, thus indicating an only incomplete splitting. The relation of our cost function to the excitation probability $p$ used in ref.~\cite{haensel:01} is simply given by $J(\psi,\lambda)\simeq\frac 12p$, assuming as usual only minor contributions from the second term in eq.~\eqref{eq:cost}.

The symbols in fig.~\ref{fig:haensch1d}(a) show that optimal control theory again allows to strongly improve the cost function. In the inset we report that for transfer times beyond say 6 ms the cost function $J$ becomes significantly lower than the control penalization $\gamma=10^{-3}$ (dotted line), and the final wavefunction $\psi(t)$ matches almost perfectly the desired groundstate wavefunction of the final double-well potential. A comparison of the optimal control $\lambda(t)$ depicted in the second inset of panel (a) with the optimized control of H\"ansel {\em et al.},\/ see inset of fig.~6 of ref.~\cite{haensel:01}, shows that both control strategies are of equal simplicity. We note that the optimal control fields of our approach perform better for very short transfer times, whereas for longer transfer times further analysis would be needed to pinpoint the advantages and disadvantages of the respective schemes.

\subsection{Magnetic confinement of Lesanovsky et al.}\label{sec:lesanovsky}

\begin{figure}
\centerline{\includegraphics[width=\columnwidth]{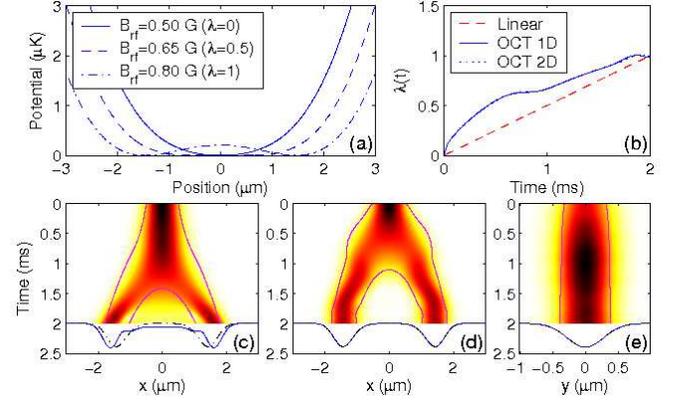}}
\caption{(Color online) Results for the magnetic microtrap of Lesanovsky {\em et al.}\/ \cite{lesanovsky:06} and for ${}^{87}$Rb atoms confined in the $m_F=2$ state. (a) Magnetic confinement potential for different rf field strengths. (b) Linear (dashed line) and optimized control parameter $\lambda(t)$ as obtained from the solutions of the one-dimensional (solid line) and two-dimensional (dotted line, indistinguishable from solid line) Schr\"odinger equation. (c) Density plot of wavefunction evolution $\int |\psi(x,y;t)|\,dy$ for linear $\lambda$ variation. (d) Same as (c) but for optimized $\lambda(t)$. (e) Density plot of wavefunction evolution $\int |\psi(x,y;t)|\,dx$ for optimized $\lambda(t)$.
}\label{fig:schmiedmayer}
\end{figure}

In our fourth case study we consider the radio-frequency double-well confinement proposed by Lesanovsky {\em et al.}\/ \cite{lesanovsky:06} which is produced by a surface-mounted dc four-wire structure on an atom chip. Such traps provide tight confinement even at large surface distances, allow for smooth potential transitions by variation of external parameters, such as rf field strengths, and are relatively robust against experimental fluctuations. In our calculations we use the same parameters as given in ref.~\cite{lesanovsky:06} (see also eq.~(10) therein), and vary the rf field strength by means of the control parameter $\lambda$ according to 
\begin{equation}
  B_{\rm rf}=0.5+\lambda\times 0.3\,\,\text{G}\,.
\end{equation}
Fig.~\ref{fig:schmiedmayer}(a) shows the confinement along $x$ for three different rf field strengths corresponding to $\lambda=0$ (solid line), $\lambda=\frac 12$ (dashed line), and $\lambda=1$ (dashed-dotted line). Contrary to the double-well potential discussed in the previous section, the magnetic confinement of Lesanovsky and coworkers exhibits a substantial extension in $y$ direction, which calls for a solution of the two-dimensional Schr\"odinger equation. In our work this is accomplished by using the split operator technique, as discussed in more detail in appendix \ref{app:details}. Figs.~\ref{fig:schmiedmayer}(b--e) show results of our optimal quantum control calculations for a quite short transfer time of $T=2$ ms. In panel (b) we report the optimized $\lambda(t)$ functions as obtained from the solutions of the one-dimensional (solid line) and two-dimensional (dotted line) Schr\"odinger equation. Both $\lambda(t)$ are almost identical. Indeed, from fig.~\ref{fig:schmiedmayer}(e), which shows the wavefunction and confinement potential along $y$, it is apparent that there is an only minor influence of $B_{\rm rf}$ on the confinement along $y$, and consequently the wavefunction factorizes. The lower parts of panels (c--e) report the final (solid line) and desired (dashed-dotted line) wavefunctions. They differ in case of a linear variation of $\lambda(t)$ [see panel (c)] and coincide for the optimized control [see panels (d,e)]. Thus, optimal quantum control allows to drastically outperform more simple control schemes. We have presented the results of fig.~\ref{fig:schmiedmayer} primarily to demonstrate our ability to also cope with two-dimensional problems. We believe that this will be important for the future analysis of more complicated potentials, such as ring-shaped interferometers \cite{lesanovsky:06}. In our concluding remarks we will further elaborate on this point.

\begin{figure}
\centerline{\includegraphics[width=0.85\columnwidth]{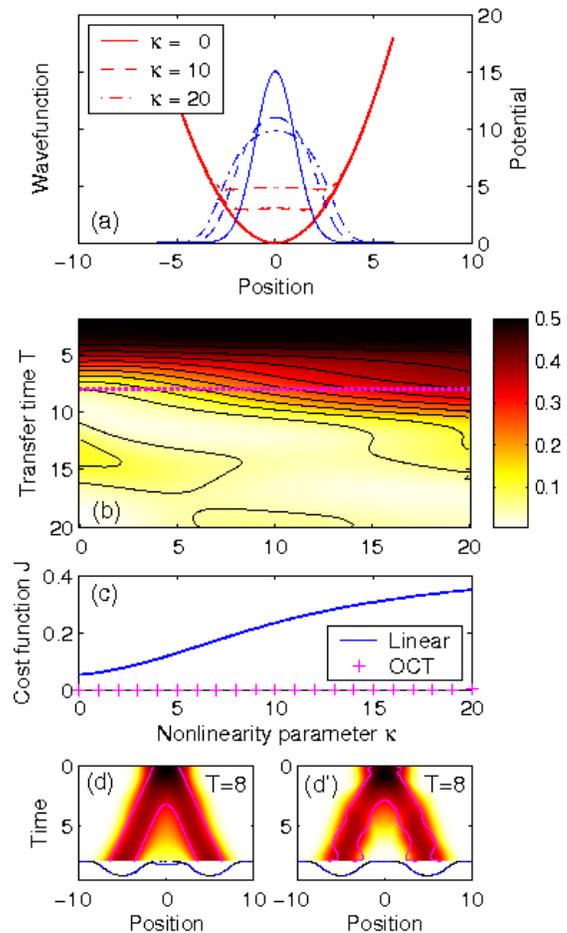}}
\caption{(Color online)
Results of our simulations performed with the nonlinear Gross--Pitaevskii equation for the double-well potential \eqref{eq:vdouble}. (a) Groundstate wavefunction $\psi_0$ and effective potential $V_0+\kappa|\psi_0|^2$ for $\lambda=0$ (single well) and different values of the nonlinearity parameter $\kappa$. (b) Density plot of the cost function $J(\psi,\lambda)$ as a function of transfer time $T$ and nonlinearity parameter $\kappa$, and for a linear $\lambda$ variation. The contour at $\kappa=0$ corresponds to the solid line shown in fig.~\ref{fig:double}(a). (c) Cost function as a function of $\kappa$ and for transfer time $T=8$. The solid line and symbols correspond to the linear and optimized variation of $\lambda$, respectively. (d,d') Density plot of $|\psi(x,t)|$ for $\kappa=20$ and $T=8$, and for a (d) linear and (d') optimized $\lambda$-variation.}\label{fig:nonlinear} 
\end{figure}

\subsection{Solution of nonlinear Gross--Pitaevskii equation}\label{sec:nonlinear}

Let us finally address the influence of the non-linear term in the Gross--Pitaevskii equation~\eqref{eq:schroedinger} on our optimal quantum control results. In doing so we shall re-examine the results of sec.~\ref{sec:double} for the simple double dot potential~\eqref{eq:vdouble}, though similar results are also found for the more realistic potentials studied in secs.~\ref{sec:haensel} and \ref{sec:lesanovsky}. Consider the one-dimensional Gross--Pitaevskii equation
\begin{equation}\label{eq:gross-pitaevskii-1d}
  i\dot\psi(x,t)=\left(-\frac 1 2 \frac{\partial^2}{\partial x^2}+V(x,\lambda(t))+
  \kappa\left|\psi(x,t)\right|^2\right)\psi(x,t)\,,
\end{equation}
where all details of the condensate density and the transversal confinement potential have been lumped into the single nonlinearity parameter $\kappa$. Figure~\ref{fig:nonlinear}(a) shows the groundstate wavefunctions $\psi_0(x)$ (see appendix~\ref{app:details-nonlinear} for computational details) and the effective potentials $V_{\rm eff}(x,\lambda)=V(x,\lambda)+\kappa|\psi_0(x)|^2$ for a few selected $\kappa$ values and for $\lambda=0$. Due to the repulsion of atoms in the condensate the wavefunction broadens and penetrates into the barrier.

We first consider condensate splitting through linear variation of $\lambda$. Fig.~\ref{fig:nonlinear}(b) shows a density plot of the corresponding cost function $J(\psi,\lambda)$ for different transfer times $T$ and nonlinearity parameters $\kappa$. Note that for $\kappa=0$ the cost function $J$ corresponds to the solid line shown in fig.~\ref{fig:double}(a). From the figure we observe that for small transfer times, say below $T=5$, the condensate becomes splitted only very inefficiently and there is no substantial overlap of the final wavefunction with the desired one. With increasing $T$ the transfer process works more efficiently. Generally speaking, for comparable values of $J$ larger nonlinearities $\kappa$ translate to longer transfer times. This is also apparent from fig.~\ref{fig:nonlinear}(c) that reports $J$ as a function of $\kappa$ for a fixed transfer time $T=8$. The symbols in the figure show results of our optimal control calculations, based on the solutions of eqs.~(\ref{eq:optimality}a--c), which demonstrate perfect condensate splitting within the whole $\kappa$ regime under consideration. The corresponding time evolutions of the control parameters $\lambda(t)$ (not shown) are similar to those shown for $\kappa=0$ in the inset of fig.~\ref{fig:double}(a). Thus, optimal control theory allows to devise efficient control strategies even in the presence of moderate condensate nonlinearities. Although the nonlinearity parameter influences the detailed time evolution of $\lambda(t)$, it has no drastic impact on the essentials and qualitative features of our findings. Similar conclusions apply to the results for other magnetic confinement potentials.

In this paper we have only considered the mean-field Gross-Pitaevskii dynamics and have neglected Bogoliubov-type quasiparticle excitations out of the condensate. This approximation is justified when the depletion of the initial Gross-Pitaevskii ground state due to quasiparticle excitations is sufficiently small \cite{leggett:01}. For given trap parameters, condensate density, and temperature, this can in principle be determined from the noncondensate normal and anomalous density matrices $\tilde n$ and $\tilde m$, respectively, to be computed from the generalized Gross-Pitaeveskii equations \cite{griffin:96,castin:98,leggett:01,morgan:04}. If $\tilde n$ and $\tilde m$ are initially negligible, they will remain negligible throughout the transfer process. Otherwise the dynamics of the Bogoliubov quasiparticles should be explicitly accounted for \cite{castin:98,morgan:04}, which, within the framework of optimal control theory, could be done by introducing additional Lagrange parameters for $\tilde n$ and $\tilde m$. Although such analysis is beyond the scope of the present paper, we expect from related studies \cite{hohenester.prl:04,hohenester.prb:06} that for this extended system OCT will not only allow to control the condensate wavefunction but also its quasiparticle excitations.

\section{Summary and discussion}\label{sec:summary}

In conclusion, we have studied within the framework of optimal control theory and the Gross-Pitaevskii equation quantum control of Bose-Einstein condensates in magnetic microtraps, which can be controlled by external parameters such as wire currents or radio-frequency fields. For a variety of magnetic confinement potentials transport and wavefunction splitting of the condensate has been analyzed, and we have demonstrated that OCT can drastically outperform more simple control strategies.

In contrast to adiabatic transfer schemes, where the control fields in the time-dependent Hamiltonian $H(t)$ must be changed sufficiently smooth, such that transitions to excited states are suppressed throughout,~\footnote{More specifically, the inequality $|\langle f|dH(t)/dt|0\rangle| \ll(E_f-E_0)^2$ must be fulfilled throughout the transfer process, with $0$ and $f$ denoting ground and excited state, respectively.} OCT allows to access excited states during the control. This opens the possibility to explore a larger portion of the control space, and enables high-fidelity quantum control even at short time scales. Furthermore, neither the wavefunctions nor energies of ground and excited states are needed in OCT calculations, which appears to be particularly advantageous for microtraps controlled by several external parameters, where the solution of the time-independent Schr\"odinger or Gross-Pitaevskii equation for every configuration of the magnetic confinement potential would be a computationally heavy task. OCT calculations for Bose Einstein condensates in magnetic microtraps are expected to be a useful and versatile tool for high-fidelity quantum control in a variety of applications.

\acknowledgements

Work supported in part by the Austrian Science Fund FWF under projet P18136--N13 and European Union Integrated Project FET/QIPC `Scala'.

\begin{appendix}

\section{Global phase}\label{app:phase}

Let 
\begin{equation}\label{eq:vtilde}
  V(\bm r,\lambda)=V_0(\lambda)+\tilde V(\bm r,\lambda)\,,
\end{equation}
where $V_0(\lambda)$ is the minimum value of the potential $V(\bm r,\lambda)$ and $\tilde V(\bm r,\lambda)$ a potential with minimum zero. Then,
\begin{equation}\label{eq:psitilde}
  \psi(t)=\exp\left(-i\phi(t)\right)\,\tilde\psi(t)
\end{equation}
is a modified wavefunction with the global phase $\phi(t)$ defined through
\begin{equation}
  \Phi(t)=\int_0^t V_0(\lambda(s))\,ds\,.
\end{equation}
Inserting wavefunction \eqref{eq:psitilde} into the Gross-Pitaevskii equation \eqref{eq:schroedinger} gives for the time derivative
\begin{equation}
  i\dot\psi=e^{-i\phi(t)}\left(i\dot{\tilde\psi}+\dot\phi(t)\,\tilde\psi\right)=
  ie^{-i\phi(t)}\dot{\tilde\psi}+V_0(\lambda(t))\,\psi\,.
\end{equation}
Here, the last term on the right-hand side is canceled by a corresponding term on the right-hand side of eq.~\eqref{eq:schroedinger}, and we obtain the modified Gross--Pitaevskii equation
\begin{equation}\label{eq:tildeschroedinger}
  i\dot{\tilde\psi}(\bm r,t)=\left(-\frac 1 2 \nabla^2+\tilde V\left(\bm r,\lambda(t)\right)+
  g\left|\tilde\psi(\bm r,t)\right|^2\right)\tilde\psi(\bm r,t)\,.
\end{equation}
It is identical to eq.~\eqref{eq:schroedinger} with the only exception that the constant contribution of the confinement potential $V(t)$ is substracted. The solution $\tilde\psi(t)$ equals $\psi(t)$ up to the global phase $\phi(t)$.

\section{Details of our numerical scheme}\label{app:details}

\subsection{Schr\"odinger equation}

In this appendix we present details of our numerical solution schemes for $\psi(t)$ and $p(t)$. Let us first neglect the nonlinear terms in eqs.~\eqref{eq:oct.forward} and \eqref{eq:oct.backward}, by setting $g=0$, and consider one spatial dimension. We discretize the time and space domain into a finite number $N_t$ and $N_x$ of subintervals of sizes $\delta t$ and $\delta x$, respectively. A discrete state variable at time $t_m$ and position $x_i$ is denoted by $\psi_i^m$. The second spatial derivative $\psi''$ is approximated by the finite difference expression $(\psi_{i+1}-2\psi_i+\psi_{i-1})/(\delta x)^2$ together with periodic boundary conditions. For the time integration of $\psi(t)$ and $p(t)$ we use the Crank-Nicholson scheme
\begin{equation}\label{eq:crank.nicholson}
  \psi^{m+1} = (\openone + i \frac{\delta t}{2} H^{m+1})^{-1}
  \, (\openone - i \frac{\delta t}{2} H^{m}) \psi^m\,.
\end{equation}
The inversion of the matrix on the right-hand side is computationally simple owing to its tridiagonal shape, which results from the above finite difference scheme. The Crank-Nicholson scheme has the advantage that the time evolution is unitary and the norm of the wavefunction is thus conserved. In our calculations we typically use values of $N_t=500$ and $N_x=500$.

For two space dimensions we again discretize the domain into equidistant subintervals and approximate the spatial derivatives by a corresponding five-point formula for the Laplacian. However, the resulting Hamiltonian matrix $H$ is no longer tridiagonal and the inversion in \eqref{eq:crank.nicholson} becomes computationally more costly. We thus employ the split-operator scheme \cite{bao:03b}. Let $H=T+V$, where $T$ is the kinetic term resulting from the discretization of the Laplacian and $V$ the magnetic confinement potential. The wavefunction at later time is then computed according to
\begin{equation}\label{eq:splitoperator}
  \psi^{m+1}=e^{-i \frac{\delta t}{2} V^{m+1}}
  e^{-i \delta t\, T}e^{-i \frac{\delta t}{2} V^{m}}\,\psi^m\,,
\end{equation}
which again is a norm-conserving scheme. Here, the action of the matrix $e^{-i \delta t\, T}$ on the wavefunction can be easily computed by means of fast Fourier transform (FFT) and its inverse. Space discretizations of typical dimension $512\times 128$ can be easily handled within such approach.

Finally, the second time derivative of $\ddot\lambda$ needed in eq.~\eqref{eq:grad.L} is approximated by the finite difference expression $(\lambda^{m+1}-2\lambda^m+\lambda^{m-1})/ (\delta t)^2$. In our calculations we set $\gamma=10^{-3}$.

\subsection{Gross--Pitaevskii equation}\label{app:details-nonlinear}

The split-operator technique can be also applied to the nonlinear Gross--Pitaevskii equation \eqref{eq:schroedinger}. We first replace in eq.~\eqref{eq:splitoperator} the potentials $V$ by the effective potentials $V_{\rm eff}=V+g|\psi|^2$. An apparent difficulty of the nonlinear term is the fact that the first exponetial on the right-hand side of eq.~\eqref{eq:splitoperator} invokes the wavefunction $\psi^{m+1}$, through the effective potential $V_{\rm eff}^{m+1}$, which is not at hand at this stage of computation. One can, however, easily check that the $e^{-i \frac{\delta t}{2} V_{\rm eff}^{m+1}}$ term only adds a phase to the different components $\psi_i^{m+1}$. Thus, the wavefunction modulus can be computed from 
\begin{equation}\label{eq:splitoperator-norm}
  \left|\psi^{m+1}\right|=\left|
  e^{-i \delta t\, T}e^{-i \frac{\delta t}{2} V_{\rm eff}^{m}}
  \,\psi^m\right|\,.
\end{equation}
Once $|\psi^{m+1}|$ is known we can determine the effective potential $V_{\rm eff}^{m+1}$, and finally compute the wavefunction at later time through
\begin{equation}\label{eq:splitoperator-nonlinear}
  \psi^{m+1}=e^{-i \frac{\delta t}{2} V_{\rm eff}^{m+1}}
  \left(e^{-i \delta t\, T}e^{-i \frac{\delta t}{2} V_{\rm eff}^{m}}
  \,\psi^m\right)\,.
\end{equation}
This scheme again conserves the norm. As for the adjoint equation \eqref{eq:oct.backward}, we use a slight variant of the split-operator technique for the linear Schr\"odinger equation, where the real and imaginary parts of the equation are separated to cope with the $g\,\psi^2p^*$ term. 
%Quite generally, for the solution of the nonlinear Gross--Pitaevskii equation somewhat larger values of $N_t$ and $N_x$ are needed to get accurate results. In our calculations we typically use values $N_t=2000$ and $N_x=1000$.

Finally, for imaginary time steps $-i\delta t$ we are able to compute the groundstate wavefunction of the Gross--Pitaevskii equation. Here, we start from the groundstate of the linear Schr\"odinger equation and evolve the system through eqs.~(\ref{eq:splitoperator-norm},\ref{eq:splitoperator-nonlinear}) in imaginary time, thus projecting out the groundstate wavefunction. After each iteration the wavefunction is normalized and the computation terminates when the wavefunction does no longer change significantly.

\section{Optimization of spatial geometries}\label{app:waveguide}

In this appendix we briefly discuss how a slight variant of the OCT scheme presented in sec.~\ref{sec:theory} would allow for an optimization of spatial geometries, such as waveguides or two-arm interferometers. The situation we have in mind is a scattering-type experiment, where the condensate, initially in state $\psi_0$, enters through a waveguide into the scattering region where it becomes split. Let us consider for simplicity a two-dimensional geometry and a condensate propagation along $x$. The objective of our optimization thus becomes the choice of the confinement potential $V(x,y)$ in the scattering region through which one can propagate the condensate without creating excitations. 

Let $\psi_d$ denote the desired outgoing state of the scattering and $V(y,\lambda(x))$ the confinement potential parameterized through the space-dependent control parameter $\lambda(x)$, with $x\in[0,L]$. Instead of eq.~\eqref{eq:cost} we introduce the cost function
\begin{equation}\label{eq:cost.spatial}
  J(\psi,\lambda)=\frac 1 2 \bigl(1-\bigl|\langle \psi_d|\psi(T)\rangle\bigr|^2\bigr)
  +\frac\gamma 2\int_0^L 
  \left(\frac{\partial\lambda(x)}{\partial x}\right)^2\,dx\,,
\end{equation}
where the last term favours a smooth spatial variation of the confinement potential $V$. Performing functional derivatives of the Lagrange function we obtain again eqs.~(\ref{eq:optimality}a,b), whereas \eqref{eq:control} has to be replaced by the space-dependent version
\begin{equation}
\gamma\frac{\partial^2\lambda(x)}{\partial x^2}=
  -\Re e 
  \left<\psi,\left(\frac{\partial V_\lambda}{\partial\lambda(x)}\right)
  p\right>\,,
  \label{eq:control.spatial}
\end{equation}
with boundary conditions $\lambda(0)=0$ and $\lambda(L)=1$. Here the expression on the right-hand side involves an integration over the transversal coordinate $y$ and time. The solution of the resulting optimality system can be performed along the same lines as for its time-dependent counterpart.

\end{appendix}


\begin{thebibliography}{45}
\expandafter\ifx\csname natexlab\endcsname\relax\def\natexlab#1{#1}\fi
\expandafter\ifx\csname bibnamefont\endcsname\relax
  \def\bibnamefont#1{#1}\fi
\expandafter\ifx\csname bibfnamefont\endcsname\relax
  \def\bibfnamefont#1{#1}\fi
\expandafter\ifx\csname citenamefont\endcsname\relax
  \def\citenamefont#1{#1}\fi
\expandafter\ifx\csname url\endcsname\relax
  \def\url#1{\texttt{#1}}\fi
\expandafter\ifx\csname urlprefix\endcsname\relax\def\urlprefix{URL }\fi
\providecommand{\bibinfo}[2]{#2}
\providecommand{\eprint}[2][]{\url{#2}}

\bibitem[{\citenamefont{Folman et~al.}(2000)\citenamefont{Folman, Kr\"uger,
  Cassettari, Hessmo, Maier, and Schmiedmayer}}]{folman:00}
\bibinfo{author}{\bibfnamefont{R.}~\bibnamefont{Folman}},
  \bibinfo{author}{\bibfnamefont{P.}~\bibnamefont{Kr\"uger}},
  \bibinfo{author}{\bibfnamefont{D.}~\bibnamefont{Cassettari}},
  \bibinfo{author}{\bibfnamefont{B.}~\bibnamefont{Hessmo}},
  \bibinfo{author}{\bibfnamefont{T.}~\bibnamefont{Maier}}, \bibnamefont{and}
  \bibinfo{author}{\bibfnamefont{J.}~\bibnamefont{Schmiedmayer}},
  \bibinfo{journal}{Phys. Rev. Lett.} \textbf{\bibinfo{volume}{84}},
  \bibinfo{pages}{4749} (\bibinfo{year}{2000}).

\bibitem[{\citenamefont{H\"ansel
  et~al.}(2001{\natexlab{a}})\citenamefont{H\"ansel, Hommelhoff, H{\"a}nsch,
  and Reichel}}]{haensel:01b}
\bibinfo{author}{\bibfnamefont{W.}~\bibnamefont{H\"ansel}},
  \bibinfo{author}{\bibfnamefont{P.}~\bibnamefont{Hommelhoff}},
  \bibinfo{author}{\bibfnamefont{T.~W.} \bibnamefont{H{\"a}nsch}},
  \bibnamefont{and} \bibinfo{author}{\bibfnamefont{J.}~\bibnamefont{Reichel}},
  \bibinfo{journal}{Nature (London)} \textbf{\bibinfo{volume}{413}},
  \bibinfo{pages}{498} (\bibinfo{year}{2001}{\natexlab{a}}).

\bibitem[{\citenamefont{Folman et~al.}(2002)\citenamefont{Folman, Kr\"uger,
  Schmiedmayer, Denschlag, and Henkel}}]{folman:02}
\bibinfo{author}{\bibfnamefont{R.}~\bibnamefont{Folman}},
  \bibinfo{author}{\bibfnamefont{P.}~\bibnamefont{Kr\"uger}},
  \bibinfo{author}{\bibfnamefont{J.}~\bibnamefont{Schmiedmayer}},
  \bibinfo{author}{\bibfnamefont{J.}~\bibnamefont{Denschlag}},
  \bibnamefont{and} \bibinfo{author}{\bibfnamefont{C.}~\bibnamefont{Henkel}},
  \bibinfo{journal}{Adv. in Atom. Mol. and Opt. Phys.}
  \textbf{\bibinfo{volume}{48}}, \bibinfo{pages}{263} (\bibinfo{year}{2002}).

\bibitem[{\citenamefont{Zoller}(2002)}]{zoller:02}
\bibinfo{author}{\bibfnamefont{P.}~\bibnamefont{Zoller}},
  \bibinfo{journal}{Nature (London)} \textbf{\bibinfo{volume}{404}},
  \bibinfo{pages}{236} (\bibinfo{year}{2002}).

\bibitem[{\citenamefont{Nielsen and Chuang}(2000)}]{nielsen:00}
\bibinfo{author}{\bibfnamefont{M.~A.} \bibnamefont{Nielsen}} \bibnamefont{and}
  \bibinfo{author}{\bibfnamefont{I.~L.} \bibnamefont{Chuang}},
  \emph{\bibinfo{title}{Quantum Computation and Quantum Information}}
  (\bibinfo{publisher}{Cambridge}, \bibinfo{address}{Cabmridge},
  \bibinfo{year}{2000}).

\bibitem[{\citenamefont{Hommelhoff et~al.}(2005)\citenamefont{Hommelhoff,
  H{\"a}sel, H{\"a}nsch, and Reichel}}]{hommelhof:05}
\bibinfo{author}{\bibfnamefont{P.}~\bibnamefont{Hommelhoff}},
  \bibinfo{author}{\bibfnamefont{W.}~\bibnamefont{H{\"a}sel}},
  \bibinfo{author}{\bibfnamefont{T.~W.} \bibnamefont{H{\"a}nsch}},
  \bibnamefont{and} \bibinfo{author}{\bibfnamefont{J.}~\bibnamefont{Reichel}},
  \bibinfo{journal}{New J. Phys.} \textbf{\bibinfo{volume}{7}},
  \bibinfo{pages}{3} (\bibinfo{year}{2005}).

\bibitem[{\citenamefont{G{\"u}nther et~al.}(2005)\citenamefont{G{\"u}nther,
  Kraft, Kemmler, Koelle, Kleiner, Zimmermann, and Fortagh}}]{guenther:05}
\bibinfo{author}{\bibfnamefont{A.}~\bibnamefont{G{\"u}nther}},
  \bibinfo{author}{\bibfnamefont{S.}~\bibnamefont{Kraft}},
  \bibinfo{author}{\bibfnamefont{M.}~\bibnamefont{Kemmler}},
  \bibinfo{author}{\bibfnamefont{D.}~\bibnamefont{Koelle}},
  \bibinfo{author}{\bibfnamefont{R.}~\bibnamefont{Kleiner}},
  \bibinfo{author}{\bibfnamefont{C.}~\bibnamefont{Zimmermann}},
  \bibnamefont{and} \bibinfo{author}{\bibfnamefont{J.}~\bibnamefont{Fortagh}},
  \bibinfo{journal}{Phys. Rev. Lett.} \textbf{\bibinfo{volume}{95}},
  \bibinfo{pages}{170405} (\bibinfo{year}{2005}).

\bibitem[{\citenamefont{Luo et~al.}(2004)\citenamefont{Luo, Kr{\"g}er, Brugger,
  Wildermuth, Gimpel, Klein, Groth, Folman, Bar-Joseph, and
  Schmiedmayer}}]{luo:04}
\bibinfo{author}{\bibfnamefont{X.}~\bibnamefont{Luo}},
  \bibinfo{author}{\bibfnamefont{P.}~\bibnamefont{Kr{\"g}er}},
  \bibinfo{author}{\bibfnamefont{K.}~\bibnamefont{Brugger}},
  \bibinfo{author}{\bibfnamefont{S.}~\bibnamefont{Wildermuth}},
  \bibinfo{author}{\bibfnamefont{H.}~\bibnamefont{Gimpel}},
  \bibinfo{author}{\bibfnamefont{M.~W.} \bibnamefont{Klein}},
  \bibinfo{author}{\bibfnamefont{S.}~\bibnamefont{Groth}},
  \bibinfo{author}{\bibfnamefont{R.}~\bibnamefont{Folman}},
  \bibinfo{author}{\bibfnamefont{I.}~\bibnamefont{Bar-Joseph}},
  \bibnamefont{and}
  \bibinfo{author}{\bibfnamefont{J.}~\bibnamefont{Schmiedmayer}},
  \bibinfo{journal}{Optics Letters} \textbf{\bibinfo{volume}{29}},
  \bibinfo{pages}{2145} (\bibinfo{year}{2004}).

\bibitem[{\citenamefont{Kr{\"u}ger et~al.}(2003)\citenamefont{Kr{\"u}ger, Luo,
  Klein, Brugger, Haase, Wildermuth, Groth, Bar-Joseph, Folman, and
  Schmiedmayer}}]{krueger:03}
\bibinfo{author}{\bibfnamefont{P.}~\bibnamefont{Kr{\"u}ger}},
  \bibinfo{author}{\bibfnamefont{X.}~\bibnamefont{Luo}},
  \bibinfo{author}{\bibfnamefont{M.~W.} \bibnamefont{Klein}},
  \bibinfo{author}{\bibfnamefont{K.}~\bibnamefont{Brugger}},
  \bibinfo{author}{\bibfnamefont{A.}~\bibnamefont{Haase}},
  \bibinfo{author}{\bibfnamefont{S.}~\bibnamefont{Wildermuth}},
  \bibinfo{author}{\bibfnamefont{S.}~\bibnamefont{Groth}},
  \bibinfo{author}{\bibfnamefont{I.}~\bibnamefont{Bar-Joseph}},
  \bibinfo{author}{\bibfnamefont{R.}~\bibnamefont{Folman}}, \bibnamefont{and}
  \bibinfo{author}{\bibfnamefont{J.}~\bibnamefont{Schmiedmayer}},
  \bibinfo{journal}{Phys. Rev. Lett.} \textbf{\bibinfo{volume}{91}},
  \bibinfo{pages}{233201} (\bibinfo{year}{2003}).

\bibitem[{\citenamefont{Schumm et~al.}(2005)\citenamefont{Schumm, Hofferberth,
  Andersson, Wildermuth, Groth, Bar-Joseph, Schmiedmayer, and
  Kr\"uger}}]{schumm:05}
\bibinfo{author}{\bibfnamefont{T.}~\bibnamefont{Schumm}},
  \bibinfo{author}{\bibfnamefont{S.}~\bibnamefont{Hofferberth}},
  \bibinfo{author}{\bibfnamefont{L.~M.} \bibnamefont{Andersson}},
  \bibinfo{author}{\bibfnamefont{S.}~\bibnamefont{Wildermuth}},
  \bibinfo{author}{\bibfnamefont{S.}~\bibnamefont{Groth}},
  \bibinfo{author}{\bibfnamefont{I.}~\bibnamefont{Bar-Joseph}},
  \bibinfo{author}{\bibfnamefont{J.}~\bibnamefont{Schmiedmayer}},
  \bibnamefont{and} \bibinfo{author}{\bibfnamefont{P.}~\bibnamefont{Kr\"uger}},
  \bibinfo{journal}{Nature Phys.} \textbf{\bibinfo{volume}{1}},
  \bibinfo{pages}{57} (\bibinfo{year}{2005}).

\bibitem[{\citenamefont{Hofferberth et~al.}(2006)\citenamefont{Hofferberth,
  Lesanovsky, Fischer, Verdu, and Schmiedmayer}}]{hofferberth:06}
\bibinfo{author}{\bibfnamefont{S.}~\bibnamefont{Hofferberth}},
  \bibinfo{author}{\bibfnamefont{I.}~\bibnamefont{Lesanovsky}},
  \bibinfo{author}{\bibfnamefont{B.}~\bibnamefont{Fischer}},
  \bibinfo{author}{\bibfnamefont{J.}~\bibnamefont{Verdu}}, \bibnamefont{and}
  \bibinfo{author}{\bibfnamefont{J.}~\bibnamefont{Schmiedmayer}},
  \bibinfo{journal}{Nature Physics} \textbf{\bibinfo{volume}{2}},
  \bibinfo{pages}{710} (\bibinfo{year}{2006}).

\bibitem[{\citenamefont{Wang et~al.}(2005)\citenamefont{Wang, Anderson, Bright,
  Cornell, Diot, Kishimoto, Prentiss, Saravanan, Segal, and Wu}}]{wang:05}
\bibinfo{author}{\bibfnamefont{Y.-J.} \bibnamefont{Wang}},
  \bibinfo{author}{\bibfnamefont{D.~Z.} \bibnamefont{Anderson}},
  \bibinfo{author}{\bibfnamefont{V.~M.} \bibnamefont{Bright}},
  \bibinfo{author}{\bibfnamefont{E.~A.} \bibnamefont{Cornell}},
  \bibinfo{author}{\bibfnamefont{Q.}~\bibnamefont{Diot}},
  \bibinfo{author}{\bibfnamefont{T.}~\bibnamefont{Kishimoto}},
  \bibinfo{author}{\bibfnamefont{M.}~\bibnamefont{Prentiss}},
  \bibinfo{author}{\bibfnamefont{R.~A.} \bibnamefont{Saravanan}},
  \bibinfo{author}{\bibfnamefont{S.~R.} \bibnamefont{Segal}}, \bibnamefont{and}
  \bibinfo{author}{\bibfnamefont{S.}~\bibnamefont{Wu}}, \bibinfo{journal}{Phys.
  Rev. Lett.} \textbf{\bibinfo{volume}{94}}, \bibinfo{pages}{090405}
  (\bibinfo{year}{2005}).

\bibitem[{\citenamefont{H\"ansel
  et~al.}(2001{\natexlab{b}})\citenamefont{H\"ansel, Reichel, Hommelhoff, and
  H{\"a}nsch}}]{haensel:01}
\bibinfo{author}{\bibfnamefont{W.}~\bibnamefont{H\"ansel}},
  \bibinfo{author}{\bibfnamefont{J.}~\bibnamefont{Reichel}},
  \bibinfo{author}{\bibfnamefont{P.}~\bibnamefont{Hommelhoff}},
  \bibnamefont{and} \bibinfo{author}{\bibfnamefont{T.~W.}
  \bibnamefont{H{\"a}nsch}}, \bibinfo{journal}{Phys. Rev. A}
  \textbf{\bibinfo{volume}{64}}, \bibinfo{pages}{063607}
  (\bibinfo{year}{2001}{\natexlab{b}}).

\bibitem[{\citenamefont{Andersson et~al.}(2002)\citenamefont{Andersson,
  Calarco, Folman, Andersson, Hessmo, and Schmiedmayer}}]{andersson:02}
\bibinfo{author}{\bibfnamefont{E.}~\bibnamefont{Andersson}},
  \bibinfo{author}{\bibfnamefont{T.}~\bibnamefont{Calarco}},
  \bibinfo{author}{\bibfnamefont{R.}~\bibnamefont{Folman}},
  \bibinfo{author}{\bibfnamefont{M.}~\bibnamefont{Andersson}},
  \bibinfo{author}{\bibfnamefont{B.}~\bibnamefont{Hessmo}}, \bibnamefont{and}
  \bibinfo{author}{\bibfnamefont{J.}~\bibnamefont{Schmiedmayer}},
  \bibinfo{journal}{Phys. Rev. Lett.} \textbf{\bibinfo{volume}{88}},
  \bibinfo{pages}{100401} (\bibinfo{year}{2002}).

\bibitem[{jo:()}]{jo:06}
\bibinfo{note}{G.-B. Jo, Y. Shin, S. Will, T. A. Pasquini, M. Saba, W.
  Ketterle, D. E. Pritchard, M. Vengalattore, and M. Prentiss, Arxiv
  cond-mat/0608585 (2006).}

\bibitem[{\citenamefont{Calarco et~al.}(2000)\citenamefont{Calarco, Hinds,
  Jaksch, Schmiedmayer, Cirac, and Zoller}}]{calarco:00}
\bibinfo{author}{\bibfnamefont{T.}~\bibnamefont{Calarco}},
  \bibinfo{author}{\bibfnamefont{E.~A.} \bibnamefont{Hinds}},
  \bibinfo{author}{\bibfnamefont{D.}~\bibnamefont{Jaksch}},
  \bibinfo{author}{\bibfnamefont{J.}~\bibnamefont{Schmiedmayer}},
  \bibinfo{author}{\bibfnamefont{J.~I.} \bibnamefont{Cirac}}, \bibnamefont{and}
  \bibinfo{author}{\bibfnamefont{P.}~\bibnamefont{Zoller}},
  \bibinfo{journal}{Phys. Rev. A} \textbf{\bibinfo{volume}{61}},
  \bibinfo{pages}{022304} (\bibinfo{year}{2000}).

\bibitem[{\citenamefont{Charron et~al.}(2006)\citenamefont{Charron, Cirone,
  Negretti, Schmiedmayer, and Calarco}}]{charron:06}
\bibinfo{author}{\bibfnamefont{E.}~\bibnamefont{Charron}},
  \bibinfo{author}{\bibfnamefont{M.}~\bibnamefont{Cirone}},
  \bibinfo{author}{\bibfnamefont{A.}~\bibnamefont{Negretti}},
  \bibinfo{author}{\bibfnamefont{J.}~\bibnamefont{Schmiedmayer}},
  \bibnamefont{and} \bibinfo{author}{\bibfnamefont{T.}~\bibnamefont{Calarco}},
  \bibinfo{journal}{Phys. Rev. A} \textbf{\bibinfo{volume}{74}},
  \bibinfo{pages}{012308} (\bibinfo{year}{2006}).

\bibitem[{\citenamefont{Treutlein et~al.}(2006)\citenamefont{Treutlein,
  H{\"a}nsch, Reichel, Negretti, Cirone, and Calarco}}]{treutlein:06}
\bibinfo{author}{\bibfnamefont{P.}~\bibnamefont{Treutlein}},
  \bibinfo{author}{\bibfnamefont{T.~W.} \bibnamefont{H{\"a}nsch}},
  \bibinfo{author}{\bibfnamefont{J.}~\bibnamefont{Reichel}},
  \bibinfo{author}{\bibfnamefont{A.}~\bibnamefont{Negretti}},
  \bibinfo{author}{\bibfnamefont{M.~A.} \bibnamefont{Cirone}},
  \bibnamefont{and} \bibinfo{author}{\bibfnamefont{T.}~\bibnamefont{Calarco}},
  \bibinfo{journal}{Phys. Rev. A} \textbf{\bibinfo{volume}{74}},
  \bibinfo{pages}{022312} (\bibinfo{year}{2006}).

\bibitem[{\citenamefont{Paul et~al.}(2005)\citenamefont{Paul, Richter, and
  Schlagheck}}]{paul:05}
\bibinfo{author}{\bibfnamefont{T.}~\bibnamefont{Paul}},
  \bibinfo{author}{\bibfnamefont{K.}~\bibnamefont{Richter}}, \bibnamefont{and}
  \bibinfo{author}{\bibfnamefont{P.}~\bibnamefont{Schlagheck}},
  \bibinfo{journal}{Phys. Rev. Lett.} \textbf{\bibinfo{volume}{94}},
  \bibinfo{pages}{020404} (\bibinfo{year}{2005}).

\bibitem[{\citenamefont{Wildermuth et~al.}(2005)\citenamefont{Wildermuth,
  Hofferberth, Lesanovsky, Haller, Mauritz-Andersson, Groth, Bar-Joseph,
  Kr{\"u}ger, and Schmiedmayer}}]{wildermuth:05}
\bibinfo{author}{\bibfnamefont{S.}~\bibnamefont{Wildermuth}},
  \bibinfo{author}{\bibfnamefont{S.}~\bibnamefont{Hofferberth}},
  \bibinfo{author}{\bibfnamefont{I.}~\bibnamefont{Lesanovsky}},
  \bibinfo{author}{\bibfnamefont{E.}~\bibnamefont{Haller}},
  \bibinfo{author}{\bibfnamefont{L.}~\bibnamefont{Mauritz-Andersson}},
  \bibinfo{author}{\bibfnamefont{S.}~\bibnamefont{Groth}},
  \bibinfo{author}{\bibfnamefont{I.}~\bibnamefont{Bar-Joseph}},
  \bibinfo{author}{\bibfnamefont{P.}~\bibnamefont{Kr{\"u}ger}},
  \bibnamefont{and}
  \bibinfo{author}{\bibfnamefont{J.}~\bibnamefont{Schmiedmayer}},
  \bibinfo{journal}{Nature (London)} \textbf{\bibinfo{volume}{435}},
  \bibinfo{pages}{440} (\bibinfo{year}{2005}).

\bibitem[{\citenamefont{Lesanovsky
  et~al.}(2006{\natexlab{a}})\citenamefont{Lesanovsky, Schumm, Hofferberth,
  Andersson, Kr\"uger, and Schmiedmayer}}]{lesanovsky:06}
\bibinfo{author}{\bibfnamefont{I.}~\bibnamefont{Lesanovsky}},
  \bibinfo{author}{\bibfnamefont{T.}~\bibnamefont{Schumm}},
  \bibinfo{author}{\bibfnamefont{S.}~\bibnamefont{Hofferberth}},
  \bibinfo{author}{\bibfnamefont{L.~M.} \bibnamefont{Andersson}},
  \bibinfo{author}{\bibfnamefont{P.}~\bibnamefont{Kr\"uger}}, \bibnamefont{and}
  \bibinfo{author}{\bibfnamefont{J.}~\bibnamefont{Schmiedmayer}},
  \bibinfo{journal}{Phys. Rev. A} \textbf{\bibinfo{volume}{73}},
  \bibinfo{eid}{033619} (\bibinfo{year}{2006}{\natexlab{a}}).

\bibitem[{\citenamefont{Lesanovsky
  et~al.}(2006{\natexlab{b}})\citenamefont{Lesanovsky, Hofferberth,
  Schmiedmayer, and Schmelcher}}]{lesanovsky:06b}
\bibinfo{author}{\bibfnamefont{I.}~\bibnamefont{Lesanovsky}},
  \bibinfo{author}{\bibfnamefont{S.}~\bibnamefont{Hofferberth}},
  \bibinfo{author}{\bibfnamefont{J.}~\bibnamefont{Schmiedmayer}},
  \bibnamefont{and}
  \bibinfo{author}{\bibfnamefont{P.}~\bibnamefont{Schmelcher}},
  \bibinfo{journal}{Phys. Rev. A} \textbf{\bibinfo{volume}{74}},
  \bibinfo{pages}{033619} (\bibinfo{year}{2006}{\natexlab{b}}).

\bibitem[{\citenamefont{Wildermuth et~al.}(2006)\citenamefont{Wildermuth,
  Hofferberth, Lesanovsky, Groth, Kr{\"u}ger, Schmiedmayer, and
  Bar-Joseph}}]{wildermuth:06}
\bibinfo{author}{\bibfnamefont{S.}~\bibnamefont{Wildermuth}},
  \bibinfo{author}{\bibfnamefont{S.}~\bibnamefont{Hofferberth}},
  \bibinfo{author}{\bibfnamefont{I.}~\bibnamefont{Lesanovsky}},
  \bibinfo{author}{\bibfnamefont{S.}~\bibnamefont{Groth}},
  \bibinfo{author}{\bibfnamefont{P.}~\bibnamefont{Kr{\"u}ger}},
  \bibinfo{author}{\bibfnamefont{J.}~\bibnamefont{Schmiedmayer}},
  \bibnamefont{and}
  \bibinfo{author}{\bibfnamefont{I.}~\bibnamefont{Bar-Joseph}},
  \bibinfo{journal}{Appl. Phys. Lett.} \textbf{\bibinfo{volume}{88}},
  \bibinfo{pages}{264103} (\bibinfo{year}{2006}).

\bibitem[{\citenamefont{Dalfovo et~al.}(1999)\citenamefont{Dalfovo, Giorgini,
  Pitaevskii, and Stringari}}]{dalfovo:99}
\bibinfo{author}{\bibfnamefont{F.}~\bibnamefont{Dalfovo}},
  \bibinfo{author}{\bibfnamefont{S.}~\bibnamefont{Giorgini}},
  \bibinfo{author}{\bibfnamefont{L.~P.} \bibnamefont{Pitaevskii}},
  \bibnamefont{and}
  \bibinfo{author}{\bibfnamefont{S.}~\bibnamefont{Stringari}},
  \bibinfo{journal}{Rev. Mod. Phys.} \textbf{\bibinfo{volume}{71}},
  \bibinfo{pages}{463} (\bibinfo{year}{1999}).

\bibitem[{\citenamefont{Leggett}(2001)}]{leggett:01}
\bibinfo{author}{\bibfnamefont{A.}~\bibnamefont{Leggett}},
  \bibinfo{journal}{Rev. Mod. Phys.} \textbf{\bibinfo{volume}{73}},
  \bibinfo{pages}{307} (\bibinfo{year}{2001}).

\bibitem[{\citenamefont{Peirce et~al.}(1988)\citenamefont{Peirce, Dahleh, and
  Rabitz}}]{peirce:88}
\bibinfo{author}{\bibfnamefont{A.~P.} \bibnamefont{Peirce}},
  \bibinfo{author}{\bibfnamefont{M.~A.} \bibnamefont{Dahleh}},
  \bibnamefont{and} \bibinfo{author}{\bibfnamefont{H.}~\bibnamefont{Rabitz}},
  \bibinfo{journal}{Phys. Rev. A} \textbf{\bibinfo{volume}{37}},
  \bibinfo{pages}{4950} (\bibinfo{year}{1988}).

\bibitem[{\citenamefont{Borz{\`\i} et~al.}(2002)\citenamefont{Borz{\`\i},
  Stadler, and Hohenester}}]{borzi.pra:02}
\bibinfo{author}{\bibfnamefont{A.}~\bibnamefont{Borz{\`\i}}},
  \bibinfo{author}{\bibfnamefont{G.}~\bibnamefont{Stadler}}, \bibnamefont{and}
  \bibinfo{author}{\bibfnamefont{U.}~\bibnamefont{Hohenester}},
  \bibinfo{journal}{Phys. Rev. A} \textbf{\bibinfo{volume}{66}},
  \bibinfo{pages}{053811} (\bibinfo{year}{2002}).

\bibitem[{\citenamefont{Rabitz et~al.}(2000)\citenamefont{Rabitz, {de
  Vivie-Riedle}, Motzkus, and Kompka}}]{rabitz:00}
\bibinfo{author}{\bibfnamefont{H.}~\bibnamefont{Rabitz}},
  \bibinfo{author}{\bibfnamefont{R.}~\bibnamefont{{de Vivie-Riedle}}},
  \bibinfo{author}{\bibfnamefont{M.}~\bibnamefont{Motzkus}}, \bibnamefont{and}
  \bibinfo{author}{\bibfnamefont{K.}~\bibnamefont{Kompka}},
  \bibinfo{journal}{Science} \textbf{\bibinfo{volume}{288}},
  \bibinfo{pages}{824} (\bibinfo{year}{2000}).

\bibitem[{\citenamefont{Tesch and {de Vivie-Riedle}}(2002)}]{tesch:02}
\bibinfo{author}{\bibfnamefont{C.~M.} \bibnamefont{Tesch}} \bibnamefont{and}
  \bibinfo{author}{\bibfnamefont{R.}~\bibnamefont{{de Vivie-Riedle}}},
  \bibinfo{journal}{Phys. Rev. Lett.} \textbf{\bibinfo{volume}{89}},
  \bibinfo{pages}{157901} (\bibinfo{year}{2002}).

\bibitem[{\citenamefont{Calarco et~al.}(2004)\citenamefont{Calarco, Dorner,
  Julienne, Williams, and Zoller}}]{calarco:04}
\bibinfo{author}{\bibfnamefont{T.}~\bibnamefont{Calarco}},
  \bibinfo{author}{\bibfnamefont{U.}~\bibnamefont{Dorner}},
  \bibinfo{author}{\bibfnamefont{P.~S.} \bibnamefont{Julienne}},
  \bibinfo{author}{\bibfnamefont{C.~J.} \bibnamefont{Williams}},
  \bibnamefont{and} \bibinfo{author}{\bibfnamefont{P.}~\bibnamefont{Zoller}},
  \bibinfo{journal}{Phys. Rev. A} \textbf{\bibinfo{volume}{70}},
  \bibinfo{pages}{012306} (\bibinfo{year}{2004}).

\bibitem[{\citenamefont{Koch et~al.}(2004)\citenamefont{Koch, Palao, Kosloff,
  and Masnou-Seeuws}}]{koch:04}
\bibinfo{author}{\bibfnamefont{C.~P.} \bibnamefont{Koch}},
  \bibinfo{author}{\bibfnamefont{J.~P.} \bibnamefont{Palao}},
  \bibinfo{author}{\bibfnamefont{R.}~\bibnamefont{Kosloff}}, \bibnamefont{and}
  \bibinfo{author}{\bibfnamefont{F.}~\bibnamefont{Masnou-Seeuws}},
  \bibinfo{journal}{Phys. Rev. A} \textbf{\bibinfo{volume}{70}},
  \bibinfo{pages}{013402} (\bibinfo{year}{2004}).

\bibitem[{\citenamefont{Hohenester and Stadler}(2004)}]{hohenester.prl:04}
\bibinfo{author}{\bibfnamefont{U.}~\bibnamefont{Hohenester}} \bibnamefont{and}
  \bibinfo{author}{\bibfnamefont{G.}~\bibnamefont{Stadler}},
  \bibinfo{journal}{Phys. Rev. Lett.} \textbf{\bibinfo{volume}{92}},
  \bibinfo{pages}{196801} (\bibinfo{year}{2004}).

\bibitem[{\citenamefont{Henkel et~al.}(1999)\citenamefont{Henkel, P\"otting,
  and Wilkens}}]{henkel:99}
\bibinfo{author}{\bibfnamefont{C.}~\bibnamefont{Henkel}},
  \bibinfo{author}{\bibfnamefont{S.}~\bibnamefont{P\"otting}},
  \bibnamefont{and} \bibinfo{author}{\bibfnamefont{M.}~\bibnamefont{Wilkens}},
  \bibinfo{journal}{Appl. Phys. B} \textbf{\bibinfo{volume}{69}},
  \bibinfo{pages}{379} (\bibinfo{year}{1999}).

\bibitem[{\citenamefont{Scheel et~al.}(2005)\citenamefont{Scheel, Rekdal,
  Knight, and Hinds}}]{scheel:05}
\bibinfo{author}{\bibfnamefont{S.}~\bibnamefont{Scheel}},
  \bibinfo{author}{\bibfnamefont{P.~K.} \bibnamefont{Rekdal}},
  \bibinfo{author}{\bibfnamefont{P.~L.} \bibnamefont{Knight}},
  \bibnamefont{and} \bibinfo{author}{\bibfnamefont{E.~A.} \bibnamefont{Hinds}},
  \bibinfo{journal}{Phys. Rev. A} \textbf{\bibinfo{volume}{72}},
  \bibinfo{pages}{042901} (\bibinfo{year}{2005}).

\bibitem[{\citenamefont{Skagerstam et~al.}(2006)\citenamefont{Skagerstam,
  Hohenester, Eiguren, and Rekdal}}]{skagerstam:06}
\bibinfo{author}{\bibfnamefont{B.~S.} \bibnamefont{Skagerstam}},
  \bibinfo{author}{\bibfnamefont{U.}~\bibnamefont{Hohenester}},
  \bibinfo{author}{\bibfnamefont{A.}~\bibnamefont{Eiguren}}, \bibnamefont{and}
  \bibinfo{author}{\bibfnamefont{P.~K.} \bibnamefont{Rekdal}},
  \bibinfo{journal}{Phys. Rev. Lett.} \textbf{\bibinfo{volume}{97}},
  \bibinfo{pages}{070401} (\bibinfo{year}{2006}).

\bibitem[{\citenamefont{Jirari and P{\"o}tz}(2005)}]{jirari:05}
\bibinfo{author}{\bibfnamefont{H.}~\bibnamefont{Jirari}} \bibnamefont{and}
  \bibinfo{author}{\bibfnamefont{W.}~\bibnamefont{P{\"o}tz}},
  \bibinfo{journal}{Phys. Rev. A} \textbf{\bibinfo{volume}{72}},
  \bibinfo{pages}{013409} (\bibinfo{year}{2005}).

\bibitem[{\citenamefont{Hohenester}(2006)}]{hohenester.prb:06}
\bibinfo{author}{\bibfnamefont{U.}~\bibnamefont{Hohenester}},
  \bibinfo{journal}{Phys. Rev. B} \textbf{\bibinfo{volume}{74}},
  \bibinfo{eid}{161307} (\bibinfo{year}{2006}).

\bibitem[{\citenamefont{Bertsekas}(1999)}]{bertsekas:99}
\bibinfo{author}{\bibfnamefont{D.~P.} \bibnamefont{Bertsekas}},
  \emph{\bibinfo{title}{Nonlinear Programming}} (\bibinfo{publisher}{Athena
  Scientific}, \bibinfo{address}{Cambridge, UK}, \bibinfo{year}{1999}).

\bibitem[{\citenamefont{Caves et~al.}(1980)\citenamefont{Caves, Thorne, Drever,
  Sandberg, and Zimmermann}}]{caves:80}
\bibinfo{author}{\bibfnamefont{C.~M.} \bibnamefont{Caves}},
  \bibinfo{author}{\bibfnamefont{K.~S.} \bibnamefont{Thorne}},
  \bibinfo{author}{\bibfnamefont{R.~W.~P.} \bibnamefont{Drever}},
  \bibinfo{author}{\bibfnamefont{V.~D.} \bibnamefont{Sandberg}},
  \bibnamefont{and}
  \bibinfo{author}{\bibfnamefont{M.}~\bibnamefont{Zimmermann}},
  \bibinfo{journal}{Rev. Mod. Phys.} \textbf{\bibinfo{volume}{52}},
  \bibinfo{pages}{341} (\bibinfo{year}{1980}).

\bibitem[{\citenamefont{Scully and Zubairy}(1997)}]{scully:97}
\bibinfo{author}{\bibfnamefont{M.~O.} \bibnamefont{Scully}} \bibnamefont{and}
  \bibinfo{author}{\bibfnamefont{M.~S.} \bibnamefont{Zubairy}},
  \emph{\bibinfo{title}{Quantum Optics}} (\bibinfo{publisher}{Cambridge
  University Press}, \bibinfo{address}{Cambridge, UK}, \bibinfo{year}{1997}).

\bibitem[{\citenamefont{Zurek}(2003)}]{zurek:03}
\bibinfo{author}{\bibfnamefont{W.~H.} \bibnamefont{Zurek}},
  \bibinfo{journal}{Rev. Mod. Phys.} \textbf{\bibinfo{volume}{75}},
  \bibinfo{pages}{715} (\bibinfo{year}{2003}).

\bibitem[{\citenamefont{Griffin}(1996)}]{griffin:96}
\bibinfo{author}{\bibfnamefont{A.}~\bibnamefont{Griffin}},
  \bibinfo{journal}{Phys. Rev. B} \textbf{\bibinfo{volume}{53}},
  \bibinfo{pages}{9341} (\bibinfo{year}{1996}).

\bibitem[{\citenamefont{Castin and Dum}(1998)}]{castin:98}
\bibinfo{author}{\bibfnamefont{Y.}~\bibnamefont{Castin}} \bibnamefont{and}
  \bibinfo{author}{\bibfnamefont{R.}~\bibnamefont{Dum}},
  \bibinfo{journal}{Phys. Rev. A} \textbf{\bibinfo{volume}{57}},
  \bibinfo{pages}{3008} (\bibinfo{year}{1998}).

\bibitem[{\citenamefont{Morgan}(2004)}]{morgan:04}
\bibinfo{author}{\bibfnamefont{S.~A.} \bibnamefont{Morgan}},
  \bibinfo{journal}{Phys. Rev. A} \textbf{\bibinfo{volume}{69}},
  \bibinfo{pages}{023609} (\bibinfo{year}{2004}).

\bibitem[{\citenamefont{Bao et~al.}(2003)\citenamefont{Bao, Jim, and
  Markovich}}]{bao:03b}
\bibinfo{author}{\bibfnamefont{W.}~\bibnamefont{Bao}},
  \bibinfo{author}{\bibfnamefont{S.}~\bibnamefont{Jim}}, \bibnamefont{and}
  \bibinfo{author}{\bibfnamefont{P.~A.} \bibnamefont{Markovich}},
  \bibinfo{journal}{SIAM J. Sci. Comput.} \textbf{\bibinfo{volume}{25}},
  \bibinfo{pages}{27} (\bibinfo{year}{2003}).

\end{thebibliography}
\end{document}